\documentclass [12pt] {article}
\usepackage{a4}
 
\usepackage{amsfonts}
\usepackage{amsmath} 
\usepackage{graphics}
\usepackage{epsfig}

\author{T.~Leonhardt, R.~Manvelyan and W.~R\"uhl}
\title{The group approach to AdS space propagators}

\begin{document}

\newcommand {\eps}{\varepsilon}
\newcommand {\ti}{\tilde}
\newcommand {\D}{\Delta}
\newcommand {\G}{\Gamma}
\newcommand {\de}{\delta}
\newcommand {\al}{\alpha}
\newcommand {\la}{\lambda}
\newcommand {\si}{\sigma}
\newcommand {\vn}{\vec \nabla}
\newcommand {\vp}{\vec \partial}
\newcommand {\La}{\Lambda}

\newtheorem{lem}{Lemma}

\thispagestyle{empty}

\noindent hep-th/0305235   \hfill  May 2003 \\                   
 
\noindent
\vskip3.3cm
\begin{center}
 
{\Large\bf The group approach to AdS space propagators }
\bigskip\bigskip\bigskip
 
{\large Thorsten Leonhardt, Ruben Manvelyan\footnote{On leave from Yerevan Physics Institute, e-mail: manvel@moon.yerphi.am} and Werner R\"uhl}
\medskip
 
{\small\it Department of Physics\\ Erwin Schr\"odinger Stra\ss e \\
University of Kaiserslautern, Postfach 3049}\\ {\small\it 67653
Kaiserslautern, Germany}\\
\medskip
{\small\tt tleon,manvel,ruehl@physik.uni-kl.de}
\end{center}
 
\bigskip 
\begin{center}
{\sc Abstract}
\end{center}
\noindent
We show that AdS two-point functions can be obtained by connecting two
points in the interior of AdS space with one point on its boundary by
a dual pair of Dobrev's boundary-to-bulk intertwiners and integrating
over the boundary point.

\newpage


\section{Introduction}

Though AdS field theory is a classical subject in field theory the
appearance of the AdS/CFT correspondence \cite{Maldacena:1997re} has
revived interest in this subject considerably. In most investigations
of AdS fields it is tacitly assumed that such a theory is based on a
lagrangian with interactions which are treated by a perturbative
expansion. In a seminal paper Fronsdal \cite{Fronsdal:1978rb} showed
that massless higher spin (tensor) fields can be defined and posess a
vanishing double trace. In a long series of articles M.~A.~Vasiliev
\cite{Vasiliev:2003ev} has studied an interacting theory of infinitely
many tensor fields of all ranks, which is invariant under a
generalized gauge symmetry and perturbative with respect to a small
curvature parameter. These interactions, including gravity, are only
possible on AdS space.

The AdS/CFT correspondence maps an AdS field theory with higher spin
gauge fields holographically on a conformal field theory on Minkowski
space containing a tensor current source for each gauge field. Several
such conformal field theories are known: The critical $O(N)$ sigma
model at large $N$ in $d$ spacetime dimensions, where $2<d<4$, or the
$\mathcal{N} = 4$ supersymmetric Yang-Mills theory in $4$ dimensions
and small `t Hooft coupling $\la$. The first class of models has been
proposed as a candidate for such a correspondence by Klebanov and
Polyakov \cite{Klebanov:2002ja}. In these models all rank $l$ currents
have been constructed by operator product expansion and their
anomalous dimensions were calculated in
\cite{Lang:1992zw,Lang:1992pp}, with the result that they are all
nonzero for tensor rank $l \not= 2$. In addition these models possess
a scalar field of approximate conformal dimension $2$, independent of
$d$, which is termed ``auxiliary'' or ``Lagrangrian multiplier''
field. Its $3$- and $4$-point functions are known
\cite{Lang:1992zw,Lang:1992pp} and serve as a source of detailed
information such as the coupling constants between two scalar fields
and a current \cite{Leonhardt:2002sn}. As a whole they offer themselves
as a test object for the AdS/CFT correspondence as proposed by
Klebanov and Polyakov. To construct an AdS field candidate for this
purpose and to test the correspondence one has to compute the
respective $3$ and $4$-point functions. As a first step in this
direction we have developed an algorithm for the derivation of
two-point functions in the AdS theory (``bulk-to-bulk propagators'')
for all traceless symmetric tensor fields. In a Lagrangian setting
they represent only one irreducible component of a tensor
field. However, whether the AdS field theory is Lagrangian or, such as
conformal field theory, based on a set of fundamental fields, a
skeleton expansion and representation theory need not be answered at
the start of the investigation. With this article we hope to put
representation theory in the right position.

Propagators of symmetric tensor fields in AdS are known for the ranks
$l=0$ \cite{Burgess:ti}, $l=1$ \cite{Allen:wd,D'Hoker:1999jc} and $l=2$ \cite{Allen:1986tt,Turyn:1988af,D'Hoker:1999jc}. They have been derived from field equations and the
requirement of a specific asymptotic behaviour on the boundary of AdS
space. The $l=2$ field has a nonvanishing trace. Our approach is
based on representation theory and the use of intertwiners constructed
by Dobrev \cite{Dobrev:1998md}, which are bulk-to-boundary
propagators. It applies to all kinds of tensor fields, but for
comparison we have treated the cases of symmetric traceless tensors of
ranks $l \in \{0,1,2\}$ explicitly.

The basic notions of representation theory are presented in section
\ref{setting}. In section \ref{integrals} we evaluate the convolution
of two scalar bulk-to-boundary propagators by integration over the
boundary. It is the crucial constructive element for any bulk-to-bulk
propagator. These integrals are first presented in the form of
Appell's $F_4$ functions of two variables. However, as proved by Allen
and Jacobson \cite{Allen:wd}, there must be a representation of these
functions in terms of the geodesic distance alone, thereby replacing
the $F_4$ function by functions of a single variable. This is proved
to be possible indeed, the resulting functions are Legendre functions
of the second kind, which are two-parameter Gaussian hypergeometric
functions. The propagators are then expressed in terms of rank $l$
monomials of basic bitensors. In an appendix the technical problem of
extracting traces from these monomials of bitensors is studied. Due to
the small number of such bitensors the extraction leads to an
overdetermined system of linear equations (for $l \ge 4$), but the
excessive equations can be shown to be linearly dependent on the
others, thereby resulting in a unique solution.


\section{The setting}\label{setting}

\subsection{Euclidean conformal field theory}

We consider a euclidean conformal field theory in $d$ space(time)
dimensions. The isometry group is then given by the ``conformal
group'' $G=SO(d+1,1)$. The fields in this theory are characterized by
their transformation behaviour given by a representation $\chi$ of
$G$, which is induced from the subgroups of euclidean rotations
$M=SO(d)$ and the behaviour under dilatations, i.e.
\begin{align}
\phi(\vec x) & \mapsto r^\D \phi(r \vec x), \quad r\in \mathbb{R}_+, \nonumber \\
\phi(\vec x) & \mapsto D^\nu(m) \phi(m^{-1}\vec x), \quad m \in M.
\end{align}
Here, $\D$ is the conformal dimension of the field and $D^\nu(m)$ is
the $ \nu $ representation matrix of $ m \in M $. This determines the
representation $ \chi $ of $G$, thus we will write $ \chi = [\nu, \D]
$. Moreover, let us denote by $ \tilde \chi = [\tilde \nu, d-\D] $ the
quantum numbers of the ``shadow field'', which is obtained by
exchanging the $M$ representation $ \nu $ by its mirror image $ \tilde
\nu $ and the conformal weight $ \D $ by $ d-\D $. Since the dimension
of the shadow field appears very often, it is convenient to introduce the
notation
\begin{align}
\la := d-\D. 
\end{align}
For generic conformal dimension $ \D $, the representations $ \chi $
and $ \tilde \chi $ are equivalent, i.e. there exists an invertible
operator 
\begin{align}
G_\chi: C_{\tilde \chi} \to C_\chi,
\end{align}
which maps the representation spaces to each other and which commutes
with the action of $G$. Such an operator is called intertwining
operator for $ C_{\tilde \chi} $ and $C_{ \chi}$. It is well known
that the two-point function of a conformal field with quantum numbers $\chi$
agrees with this intertwiner $G_\chi$. The functional form of a
two-point function with quantum numbers $\chi$ itself is fixed by conformal
covariance to 
\begin{align}
G_{\chi} (\vec x) & = \frac{\gamma_\chi}{(\vec x^2)^\D} D^\nu (r(\vec
x)) \nonumber \\
 r(\vec x) & = \Bigl( \frac{2 x_i x_j}{\vec x^2} -
\delta_{ij} \Bigr), \quad i,j = 1,\ldots,d.
\end{align} 
In this formula $\gamma_\chi$ is a normalization constant, whose value
is not important for our purposes.  For later use, let us write down
the explicit form of a propagator of a symmetric traceless tensor
field $t^{(l)} $ of conformal dimension $\D$. In this case we can
write down a generating function:
\begin{align}
(\vec x^2)^\D \sum_{l\ge0} \rho^l \sum_{(\tau)(\sigma)} \langle
t^{(l)}_{(\tau)} (\vec x) t^{(l)}_{(\sigma)} (0) \rangle \vec
a^{\otimes l}_{(\tau)} \vec b^{\otimes l}_{(\sigma)} & = \Bigl( 1-2
\rho (\vec a, r(\vec x) \vec b) +\rho^2 \vec a^2 \vec b^2
\Bigr)^{1-\mu} \nonumber \\
 & = \sum_{l\ge0} \rho^l (|\vec a| |\vec
b|)^l C_l^{\mu-1} \Bigl( \frac{(\vec a, r(\vec x) \vec b)}{(|\vec a|
|\vec b|)} \Bigr), 
\end{align}
where $\vec a, \vec b \in \mathbb{R}^d$ and $C_l^{\mu-1}$ is the
Gegenbauer polynomial of degree $l$ with parameter $\mu-1$. Here we
use the convenient abbreviation $\mu= d/2$, which will be applied
throughout this work.

\subsection{AdS/CFT correspondence as representation equivalence}

On the other hand we consider a field theory on a $d+1$-dimensional
Anti-de-Sitter space, which is given by a classical action. The
AdS/CFT correspondence connects this field theory to a theory living
on the boundary of the AdS space. The isometry group of
$d+1$-dimensional AdS space is $G=SO(d+1,1)$, thus matching the
isometry groups of both theories implies that the boundary theory must
be conformal. By an identification of generating functionals we obtain
a prescription for constructing $d$-dimensional correlation functions,
which in turn define a conformal field theory. The procedure may be
sketched as follows: We take a set of vertices in the AdS theory,
connect some of them by bulk-to-bulk propagators and the remaining
vertices to the boundary by bulk-to-boundary propagators
\cite{Witten:1998qj}.  These two kinds of propagators can be obtained
by solving the free equations of motion derived from the action of the
AdS theory.  Let us write down the actual form of the scalar
bulk-to-boundary operator, which constructs an AdS scalar field of
mass $m$ from a conformal scalar of dimension $\D$. The mass $m$ is
linked to $\D$ by $m^2=\D(\D-d)$:
\begin{align}
K_\D (x_1, \vec x_2) = \Bigl(\frac{x_{10}}{ x_{12}^2}
\Bigr)^\la,
\end{align}
where $x_1 =(x_{10}, \vec x_1) \in \mathbb{R}_+ \times \mathbb{R}^d$,
$\vec x_2 \in \mathbb{R}^d$, and we introduced the notation
\begin{align}
x_{12}^2 = |x_1-\vec x_2|^2 = x_{10}^2 + (\vec x_1 - \vec x_2)^2.
\end{align}
Moreover, we note the ``vectorial'' bulk-to-boundary propagator, which
transfers a $d$-dimensional vector field of conformal dimension $\D$ to
a $d+1$-dimensional AdS-vector field of mass $m$
\begin{align} \label{vecprop}
 K^{(1)}_\D (a, \vec b; x_1, \vec x_2) = \frac{x_{10}^{\la-1}}
 {(x_{12}^2)^\la}
\Bigl( \frac{2 \langle a,x_{12} \rangle (\vec b, \vec x_{12})}
{x_{12}^2} -(\vec a, \vec b) \Bigr)
\end{align}
with the relation
\begin{align}
m^2 = \D(\D-d) + (d-1).
\end{align}
In (\ref{vecprop}) we contracted the external indices of the $d+1$-
and $d$-dimensional vectors with $a\in \mathbb{R}^{d+1}$ and $\vec b
\in \mathbb{R}^d$, respectively, via the $d+1$- and $d$-dimensional
standard scalar products.  The construction of the bulk-to-boundary
propagator, which connects a symmetric traceless second rank tensor
field of conformal dimension $\D$ with a symmetric traceless
second rank tensor field on $d+1$-dimensional AdS is simple and can
easily be generalized to any rank:
\begin{multline} 
K^{(2)}_\D (a, \vec b; x_1, \vec x_2) = \Bigl( K^{(1)}_{\la/2} (a,
\vec b; x_1, \vec x_2)\Bigr)^2 - \frac{\vec b^2}{d} \times
\textrm{trace with respect to }\vec b \\ =
\frac{x_{10}^{\la-2}}{(x_{12}^2)^{\la}} \biggl\{ \Bigl( \frac{2
\langle a,x_{12} \rangle (\vec x_{12}, \vec b)} {x_{12}^2} -(\vec a,
\vec b) \Bigr)^2 - \frac{\vec b^2}{d}  \Bigl( \frac{2 \langle
a,x_{12} \rangle \vec x_{12}} {x_{12}^2} - \vec a \Bigr)^2
\biggr\},
\end{multline}
where we contracted the external indices with $a\in
\mathbb{R}^{d+1}$ and $\vec b \in \mathbb{R}^d$, as in
(\ref{vecprop}). After a quick computation one agrees that the trace
of this bulk-to-boundary propagator with respect to $a$ vanishes, too.

As shown by Dobrev \cite{Dobrev:1998md}, these bulk-to-boundary
operators can be given an interpretation as intertwining
operators, and they can even be constructed by this property. They map
the conformal representations $C_\chi$ to representations $\hat
C^\tau$ induced from the maximal compact subgroup $K=SO(d+1) \subset
G$, where $\tau$ is an irreducible representation of $K$ containing
the irrep $\nu$ from $\chi=[\nu, \D]$ of $M$. Now $\hat C^\tau$ is
neither uniquely determined (there are infinitely many different
irreps of $K$ containing $\nu$) nor irreducible. The lack of
uniqueness can be cured by the choice of some ``minimal'' irrep of
$K$. To obtain irreducibility, one has to impose a constraint on the
behaviour of the functions in $\hat C^\tau$ for $x_0 \to 0$. The
resulting representation $\hat C^\tau_\chi$ turns out to be
irreducible for generic $\D$, and the bulk-to-boundary operators are
then the integral kernels of the intertwiners
\begin{align}
K^\tau_\chi : C_\chi \to \hat C^\tau_\chi.
\end{align}
 Dobrev's group theoretical arguments apply for all
elementary irreps $\chi$ of $G$ and all irreps $\tau$ induced from
$K$, therefore they do not depend on any actions or field equations.

After all these preliminaries, let us describe the idea of
constructing AdS two-point functions, where we restrict on symmetric
tensor representations of $M$ and $K$, so that the respective
progagators are labeled by their tensor rank $l$ and the parameter
$\D$. It is simple to convolute the conformal intertwiner
$G^{(l)}_{d-\D}$ with the bulk-to-boundary operator $K^{(l)}_\D$ and
check that we obtain a bulk-to-boundary propagator for a conformal
field of dimension $d-\D$ \footnote{This is already proved in general
in \cite{Dobrev:1998md}}. This begs the following question: If we take
a further bulk-to-boundary propagator $K^{(l)}_\D$ of dimension $\D$ and
convolute it with the remainig $d$-dimensional leg of the resulting
$K_{d-\D}$, do we obtain a scalar bulk-to-bulk propagator
$W^{(l)}_\D$? We show by explicit construction that this guess is
almost correct, except that a certain doubling occurs. Each two-point
function $W^{(l)}_\D$ in AdS obtained this way comes along with its
shadow partner $W^{(l)}_{d-\D}$. This seems natural from the conformal
point of view, since in any CFT exchange the shadow of any exchanged
field appears, because the shadow representations are equivalent to
the original ones.


\subsection{Geometric properties of AdS propagators}

Let us mention the following geometric aspects of the two-point
functions: The two-point function $\langle t^{(l)}(x) t^{(l)}(x')
\rangle$ transforms under a coordinate transformation as a tensor of
rank $l$ at the points $x$ and $x'$, i.e. the two-point function is a
``bitensor''. Thanks to the maximal symmetry of the (Euclidean) AdS
space we have the theorem \cite{Allen:wd} that every bitensor can be
expressed in terms of the metric and three fundamental bitensors,
which are obtained by differentiating the geodesic distance
$\mu(x,x')$ of two points $x,x' \in$AdS. The first two fundamental
bitensors are the two tangent unit vectors along the geodesic
\begin{align}
n_\nu = D_\nu \mu(x,x') \quad \textrm{and} \quad n_{\nu'} =
D_{\nu'} \mu(x,x'), 
\end{align}
where the unprimed/primed indices denote tangent space indices at $x$
and $x'$, respectively, and $D$ denotes covariant derivatives. The third
fundamental bitensor is the parallel transporter along the geodesic
connecting $x$ and $x'$. We choose to work in Poincar\'e coordinates
$x=(x_0,x_1,\ldots, x_d) = (x_0,\vec x) \in \mathbb{R}_+ \times
\mathbb{R}^{d}$ for the AdS space, in which the geodesic distance and
the fundamental bitensors are expressed as 
\begin{align} \label{fundbitens}
\zeta & = \cosh(\mu(x,x')) = \frac{x_0^2 + {x_0'}^2 + (\vec x - \vec
x')^2}{2 x_0 x_0'}, \nonumber \\
n_\nu & = \frac{\partial_\nu \zeta}{\sqrt{\zeta^2-1}},
\nonumber \\
g_{\nu,\nu'} & = \partial_\nu \partial_{\nu'} \zeta +
\frac{\partial_\nu \zeta \partial_{\nu'} \zeta}{\zeta+1}.
\end{align}
We note that the tangent vectors are proportional to the first
derivative of $\zeta$ and the parallel transporter is essentially
given by two derivatives of $\zeta$, one with respect to each
variable. Moreover, it turns out to be convenient to contract these
bitensors with tangent vectors $a\in T_x$AdS, $c\in T_{x'}$AdS; we
denote this by the $d+1$-dimensional scalar product $\langle \cdots
\rangle$. Then we arrive at the following algebraic basis of maximally
symmetric bitensors
\begin{align}\label{Invars1}
I_1 & := \langle a,\partial\rangle \langle c,\partial' \rangle \zeta
\nonumber \\
I_{2a} &:= \langle a, \partial \zeta \rangle \nonumber \\
I_{2c} &:= \langle c, \partial' \zeta \rangle \nonumber \\
I_2 & := I_{2a} I_{2b} \nonumber \\
\textrm{and} \; a^2&= \sum_{\nu=0}^d a_\nu^2 ,\quad c^2 = \sum_{\nu=0}^d c_\nu^2 \quad
\textrm{for} \;l\ge2.
\end{align}

In our construction we have to perform the splitting
$(a_0,a_1,\ldots,a_d)=(a_0,\vec a)$ and the same for $c$. The
invariants then acquire the form
\begin{align}\label{covtononcov}
\langle a,\partial\rangle \langle c,\partial' \rangle \zeta & =
- \frac{(\vec a, \vec c)} {x_0 x'_0} -\frac{a_0 c_0}{x_0 x'_0} \zeta -
\frac{c_0}{x'_0} \langle a, \partial \zeta \rangle - \frac{a_0}{x_0}
\langle c, \partial' \zeta \rangle, \\
\langle a, \partial \zeta \rangle & = (\vec a, \vp \zeta)
+a_0 \bigl( \frac{1}{x_0'} - \frac{\zeta}{x_0} \bigr), \\
\langle c, \partial' \zeta \rangle & = (\vec c, \vp' \zeta)
+c_0 \bigl( \frac{1}{x_0} - \frac{\zeta}{x_0'} \bigr).
\end{align}
We are interested in propagators for symmetric traceless tensors of
rank $l$. In order to subtract traces we need further bitensors 
\begin{align}\label{Invars2}
I_3 & := \frac{a^2}{x_0^2} I_{2c}^2 + \frac{c^2}{{x'_0}^2} I_{2a}^2
\nonumber \\
I_4 &:= \frac{a^2 c^2}{x_0^2 {x'_0}^2}.
\end{align}
Since taking traces on products of invariant bitensors (\ref{Invars1})
results in products of invariants with at least one factor $I_3$ or
$I_4$, we call the latter ones ``trace terms''.  For a symmetric
traceless tensor of rank l we obtain a basis labelled by the pairs
$\{(l_1,l_2\}$ with $l_1+l_2=l$:
\begin{align}\label{AnsFuerSpurLos}
I_1^{l_1} I_2^{l_2} - \sum A_{m_1 m_2 n_1 n_2}^{(l_1,l_2)} I_1^{m_1}
I_2^{m_2} I_3^{n_1} I_4^{n_2},
\end{align}
where the sum is restricted to 
\begin{align}
m_1+m_2+2(n_1+n_2)=l, \quad n_1+n_2>0.
\end{align}
The coefficients $A_{m_1 m_2 n_1 n_2}^{(l_1,l_2)}$ can be determined
by requiring (\ref{AnsFuerSpurLos}) to be harmonic with respect to the
(naive) Laplacian $\D_a=\sum \frac{\partial}{\partial a_\mu}
\frac{\partial}{\partial a_\mu}$. The solution
of this removing of traces will be presented in the appendix.


\section{Convolution integrals as hypergeometric
functions}\label{integrals} 

Now we do the actual computations. First we calculate the integral
\begin{align}\label{def_A}
A_{\al_1,\al_2} (x_1,x_3):=\int d^dx_2
\Bigl(\frac{x_{10}}{x_{10}^2+|\vec x_{12}|^2} \Bigr)^{\al_1}
\Bigl(\frac{x_{30}}{x_{30}^2+|\vec x_{32}|^2} \Bigr)^{\al_2}, 
\end{align}
where $\al_1, \al_2$ are two real parameters, which are chosen in such
a way to ensure convergence of the integral but are independent
otherwise. After introducing a Feynman parameter for the two
denominators the resulting $d$-dimensional integral is easy and we get
\begin{multline}\label{def_B}
A_{\al_1,\al_2} (x_1,x_3) = \pi^\mu \frac{\G(\al_1+\al_2-\mu)}
{\G(\al_1) \G(\al_2)}\, x_{10}^{\al_1}\, x_{30}^{\al_2} \int_0^1 dt \;
t^{\al_1-1} (1-t)^{\al_2-1}\\ 
\Bigl[ t(1-t) \vec x_{13}^2 + t x_{10}^2 +(1-t) x_{30}^2
\Bigl]^{-(\al_1+\al_2-\mu)}. 
\end{multline}
Now we extract $t x_{10}^2 +(1-t) x_{30}^2$ out of $[\cdots]$ in
(\ref{def_B}) and choose the coordinates $x_1, x_3$ in such a way that
the resulting expression in $[\cdots]$ can be expanded in a binomial
series. The case of arbitrary values of the coordinates are afterwards
obtained by analytical continuation. Then the Feynman parameter can be
integrated and we obtain
\begin{multline}
A_{\al_1,\al_2} (x_1,x_3) = \pi^\mu \frac{\G(\al_1+\al_2-\mu)}
{\G(\al_1) \G(\al_2)}\, x_{10}^{\al_1}\, x_{30}^{d-2\al_1 -\al_2}
\sum_{k\ge0} \frac{(\al_1+\al_2-\mu)_k}{k!} \\
\frac{\G(\al_1 + k) \G(\al_2+k)}{\G(\al_1+\al_2+2k)} 
\; \si^{-k} F \Bigl[ {\al_1 + \al_2 - \mu + k, \al_1+k \atop \al_1 +
\al_2 + 2k }  ; 1-\frac{1}{\rho} \Bigr],
\end{multline}
where we introduced the abbreviations
\begin{align}
\si := - \frac{\vec x_{13}^2}{x_{10}^2}, \quad \rho :=
\frac{x_{30}^2}{x_{10}^2}.
\end{align}
Next we apply an analytical continuation formula for the gaussian
hypergeometric function (formula 9.132,1 of \cite{GR}) to transform
the argument $1-\rho^{-1}$ to $\rho$ and obtain as a result two
hypergeometric functions:
\begin{align} \label{A_weiter}
& A_{\al_1,\al_2} (x_1,x_3) = \frac{\pi^\mu} {\G(\al_1) \G(\al_2)} \;
\frac{x_{30}^{d-\al_2}}{x_{10}^{\al_1}} \; \sum_{k \ge0}
\frac{\si^k}{k!} \nonumber \\
 & \; \biggl\{ \rho^{\al_2-\mu}
\G(\mu-\al_2) \G(\al_1+\al_2-\mu+k) \frac{\G(\al_2+k)}{\G(\mu+k)} F
\Bigl[ {\al_1+\al_2 -\mu + k, \al_2+k \atop \al_2-\mu + 1 } ; \rho
\Bigr] \nonumber \\
 & \qquad \qquad \qquad \qquad \qquad \qquad +
\G(\al_2-\mu) \G(\al_1+k) F \Bigl[ {\al_1 + k, \mu+k \atop 1+\mu -
\al_2 } ; \rho \Bigr] \Biggr\}.
\end{align}
The first term will be called ``direct term'' and the second will be
called ``shadow term''. 
 
 In the sequel all appearing propagators
can be expressed as linear combinations of $A_{\al_1,\al_2} $ and
derivatives thereof. In these applications the two parameters $\al_1$
and $\al_2$ are not independent but fulfil an equation 
\begin{align}
\al_1 + \al_2 = d + q,
\end{align}
where $q$ is an integer. One can check that it suffices to use only
one of the two summands in $A_{\al_1,\al_2}$, because the other one is
obtained by substituting the parameter $\la$ by the shadow parameter
$\D$. Therefore we insert  
\begin{align}
\al_1 = \la-r, \quad \al_2 =\D-s,\quad \textrm{with } \la+\D=d
\textrm{ and } r,s \in \mathbb{Z},
\end{align}
project onto the (say) first term in (\ref{A_weiter}) and thus define
\begin{align}\label{DefofPhi}
\Phi_{r,s}(x_1,x_3) :&=\int d^d x_2 \Bigl(
\frac{x_{10}}{x_{10}^2+|\vec x_{12}|^2} \Bigr)^{\la-r}
\Bigl(\frac{x_{30}}{x_{30}^2+|\vec x_{32}|^2} \Bigr)^{\D-s}\biggr
\rvert_{\textrm{direct term}}\nonumber \\ 
&= \frac{\pi^\mu \G(\mu-\D+s)} {\G(\la-r) \G(\D-s)} \;
\frac{x_{30}^{d-(\D-s)}}{x_{10}^{\la-r}} \rho^{\D-s-\mu} \; \sum_{k
\ge0} \frac{\si^k}{k!} \G(\mu-r-s+k) \nonumber \\
& \qquad \qquad \qquad \frac{\G(\D-s+k)}{\G(\mu+k)} F \Bigl[ {\mu
-r-s+ k, \D-s+k \atop \D-s-\mu + 1 } ; \rho \Bigr] .
\end{align}
The series of hypergeometric functions may be summed up in terms of
Appell's \mbox{$F_4$-function} (see 9.18 of \cite{GR} and references
therein):
\begin{multline}\label{Appelreihe}
\Phi_{r,s}(x_1,x_3)  = \frac{\pi^\mu \G(\mu-\D+s)} {\G(\la-r)
\G(\D-s)} \; \rho^{\frac{\D-s}{2}}  \\
\frac{\G(\mu-r-s) \G(\D-s)}{\G(\mu)} F_4( \D-s,
\mu-r-s, \D-s-\mu+1, \mu; \rho, \si)
\end{multline}
Let us note the action of a $d$-dimensional Laplacian on $\Phi_{r,s}$
\begin{multline}
(\vp,\vp) \Phi_{r,s} = \frac{\pi^\mu \G(\mu-\D+s)} {\G(\la-r)
\G(\D-s)} \frac{x_{30}^{d-(\D-s)}}{x_{10}^{\la-r}} \rho^{\D-s-\mu}
\Bigl(- \frac{4}{x_{10}^2} \Bigr) \; \sum_{k \ge0} \frac{\si^k}{k!} 
\G(\mu-r-s+k+1)\\ 
\frac{\G(\D-s+k+1)}{\G(\mu+k)} F \Bigl[ {\mu -r-s+ k+1, \D-s+k+1 \atop
\D-s-\mu + 1 }  ; \rho \Bigr],
\end{multline}
where $\vp = ( \frac{\partial}{\partial x_{1,i}} ),\;
i=1,\ldots,d$. This can be written again in terms of $\Phi_{r,s}$
\begin{align}\label{LaplaceaufPhi}
(\vp,\vp) \Phi_{r,s} = \frac{4}{x_{30}} (\D-s) (\D-s-\mu+1)
\Phi_{r,s-1} - 4 (\D-s)_2 \Phi_{r,s-2} .
\end{align}

We introduce Legendre functions of the second kind and write them in
terms of gaussian hypergeometric functions
\begin{align}\label{DefofLambda}
\La_{s,t} (\zeta):= \G(\D-s) 2^{-(\D-s)} \zeta^{-(\D-s)} F \Bigl[
{\frac{\D-s}{2}, \frac{\D-s+1}{2} \atop \D-s+t-\mu + 1 }  ; \zeta^{-2}
\Bigr] ,
\end{align}
where $s,t \in \mathbb{Z}$ and the AdS invariant variable $\zeta$ is
defined by (\ref{fundbitens}) with $x=x_1$ and $x'=x_3$.
We note the derivative of $\La_{r,s}$ with respect to $\zeta$.
\begin{align}\label{AblvonLa}
\frac{d}{d \zeta} \La_{s,t} (\zeta) & = - (\D-s) \G(\D-s) 2^{-(\D-s)}
\zeta^{-(\D-s+1)} F \Bigl[ {\frac{\D-s+1}{2}, \frac{\D-s+2}{2} \atop
\D-s+t-\mu + 1 }  ; \zeta^{-2} \Bigr] \nonumber \\ 
& = -2 \La_{s-1,t-1} (\zeta).
\end{align} 
Using this equation we get for the action of the $d$ dimensional
Laplacian $(\vp,\vp)$
\begin{align}
(\vp,\vp) \,\La_{s,t} (\zeta) = 4 \,
\La_{s-2,t-2} (\zeta)\, (\vp\zeta,\vp \zeta) - 2 \, \La_{s-1,t-1}
(\zeta)\, (\vp,\vp) \zeta.
\end{align}
Moreover, we note the following two identities, which are simple
consequences of eqns. 9.137, 6 and 12 in \cite{GR}, respectively
\begin{align}\label{Lambdaformeln}
\La_{s,t}(\zeta) & = \La_{s,t+1}(\zeta) + \frac{1}{(\D-\mu-s+t+1)_2}
\,\La_{s-2,t}(\zeta), \nonumber \\
\zeta \La_{s-1,t-1}(\zeta) & = \frac{\D-s}{2} \,\La_{s,t}(\zeta)
+\frac{1}{\Delta-\mu-s+t+1} \, \La_{s-2,t-1}(\zeta).
\end{align}

Now we want to express certain linear combinations and derivatives of
the functions $\Phi_{r,s}$ in terms of the $\La_{s,t}$. This can be
established with the help of the following two formulae. They both
hold in the case $r+s=m$:
\begin{align}\label{Formel1}
\sum_{j=0}^m \Bigl({m \atop j}\Bigr) \frac{(\la-r)_j}{(\D-s-j)_j}
\frac{1}{x_{30}^{m-j} x_{10}^j} \Phi_{r-j,s+j} (x_1,x_3) = \frac{ \pi^\mu
\G(\mu-\D+s)} {\G(\la-r) \G(\D-s)} \, \La_{s+m,m}(\zeta)
\end{align}
and
\begin{multline}\label{Formel2}
(\vp,\vp)^m \Phi_{r,s} (x_1,x_3) \\ = \frac{\pi^\mu \G(\mu-\D+s)}
{\G(\la-r) \G(\D-s)} \Bigl(\frac{-4}{x_{30}} \Bigr)^m \sum_{j=0}^m
\Bigl({m \atop j} \Bigr) \frac{(-1)^j \rho^{j/2}} {(\D -\mu-s+1)_j} \,
\La_{s-m-j,-m} (\zeta).
\end{multline}
The proof of (\ref{Formel1}) proceeds by induction, so let $m=0$, to
formulate the start of the induction. We have $r=-s$, and in this
case the \mbox{$F_4$-function} of (\ref{Appelreihe}) can be summed up
in terms of a gaussian hypergeometric function (eq. 9.182,8 of
\cite{GR}), which after some algebra with the coordinates reads:
\begin{align}\label{startofinduc}
\Phi_{r,s}(x_1,x_3) & =  \frac{\pi^\mu \G(\mu-\D+s)} {\G(\la-r)
\G(\D-s)} \; \rho^{\frac{\D-s}{2}} \nonumber \\ & \qquad \qquad
\G(\D-s) F_4( \D-s,\mu, \D-s-\mu+1, \mu; \rho, \si) \nonumber \\
& = \frac{\pi^\mu \G(\mu-\D+s)} {\G(\la-r) \G(\D-s)} \La_{s,0}
(\zeta),
\end{align}
thus showing the start of the induction of the proof of our
formula. For the inductive step, we take a look at the left hand side
of (\ref{Formel1}) for $m+1$ and use $\binom{m+1}{j} = \binom{m}{j-1}
+\binom{m}{j}$ to obtain
\begin{align}\label{zwischinproof1}
\sum_{j=0}^{m+1}&  \biggl[\Bigl({m \atop j-1}\Bigr) + \Bigl({m \atop
j}\Bigr) \biggr] \frac{(\la-r)_j}{(\D-s-j)_j}
\frac{1}{x_{30}^{m+1-j} x_{10}^j} \Phi_{r-j,s+j} (x_1,x_3) \nonumber
\\
& = \sum_{j=0}^{m} \Bigl({m \atop j}\Bigr)
\frac{(\la-r)_j}{(s-\D+1)_j}
\frac{(-1)^j}{x_{30}^{m-j} x_{10}^j} \nonumber \\
& \qquad \biggl[ -\frac{\la-r+j}{s-\D+j+1}\frac{1}{x_{10}}
\Phi_{r-j-1,s+j+1}(x_1,x_3) + \frac{1}{x_{30}} \Phi_{r-j,s+j}(x_1,x_3)
\biggr]. 
\end{align}
With formula 9.137,17 of \cite{GR}, the first summand in $[\cdots]$ of
(\ref{zwischinproof1}) can be written as
\begin{align}\label{zwischinproof}
& -\frac{\la-r+j}{s-\D+j+1} \frac{x_{30}^{\la+s+j+1}}
{x_{10}^{\la-r+j+2}} \rho^{\D-\mu-s-j-1} \frac{\pi^\mu
\G(\mu-\D+s+j+1)} {\G(\la-r+j+1) \G(\D-s-j-1)}\nonumber \\
& \qquad \sum_{k\ge 0} \si^k
\frac{\G(\mu-r-s+k) \G(\D-s-j-1+k)} {k!\, \G(\mu+k)} \nonumber \\
& \qquad \qquad \qquad\qquad \qquad \qquad F\Bigl[{\mu-r-s+k,
\D-s-j-1+k \atop \D-\mu-s-j }; \rho \Bigr]\nonumber \\
& = \frac{\pi^\mu \G(\mu-\D+s+j)} {\G(\la-r+j) \G(\D-s-j)}
\frac{x_{30}^{\la+s+j-1}} {x_{10}^{\la-r+j}} \rho^{\D-\mu-s-j}
\sum_{k\ge 0} \si^k \frac{\G(\mu-r-s+k) } {k!\, \G(\mu+k)} \nonumber
\\ 
& \qquad \G(\D-s-j+k) \biggl\{ - F \Bigl[{\mu-r-s+k, \D-s-j+k \atop
\D-\mu-s-j+1}; \rho \Bigr] \nonumber \\
& \qquad \qquad \qquad + \frac{\mu-1+k}{\D-s-j-1+k} F
\Bigl[{\mu-r-s+k, \D-s-j-1+k \atop \D-\mu-s-j+1}; \rho \Bigr] \biggr\} 
\end{align} 
The first term in $\{\cdots\}$ in (\ref{zwischinproof}) together with
the prefactors and the sum cancels the second term in $[\cdots]$ in
(\ref{zwischinproof1}), therefore we obtain
\begin{align}
\textrm{LHS of (\ref{Formel1})}\Bigr\rvert_{m+1} = \sum_{j=0}^{m}
\Bigl({m \atop j}\Bigr) \frac{(\la'-r')_j}{(s-\D')_j} 
\frac{(-1)^j}{x_{30}^{m-j} x_{10}^j} \frac{\pi}{\D'-s-j}
\Phi_{r'-j, s+j}(x_1,x_3)\Bigr\rvert_{\D',\la',\mu'},
\end{align}
where we have set $\D'=\D-1, \la'=\la-1, \mu' = \mu-1$ and $r'=r-1$,
and the function $\Phi$ is to be understood with the unprimed
parameters replaced by the primed ones. Now we recognize that $r'+s=m$, thus
the sum can by done by induction hypothesis, giving
\begin{align}
\textrm{LHS of (\ref{Formel1})}\Bigr\rvert_{m+1} & =
\frac{-\pi}{s-\D'} \frac{\pi^{\mu'} \G(\mu'-\D'+s)} {\G(\la'-r')
\G(\D'-s)}\, \La_{s+m,m}(\zeta)\Bigr\rvert_{\D',\la',\mu'} \nonumber
\\
&= \frac{\pi^\mu \G(\mu-\D+s)} {\G(\la-r) \G(\D-s)}\,
\La_{s+m+1,m+1}(\zeta),
\end{align}
where we used the definition (\ref{DefofLambda}) of $\La_{s,t}$. This
proves (\ref{Formel1}).

Using (\ref{LaplaceaufPhi}), the proof of (\ref{Formel2}) can also be
done by induction, where the start of the induction is the same as for
(\ref{Formel1}), but in this case the direct proof is even simpler. To
this end we apply $m$ times the Laplacian $(\vp,\vp)$ on the series in
(\ref{DefofPhi})
\begin{multline}
(\vp,\vp)^m \Phi_{r,s} (x_1,x_3) = \frac{\pi^\mu
\G(\mu-\D+s)}{\G(\la-r)
\G(\D-r)} \frac{x_{30}^{\la+s}} {x_{10}^{\la-r}} \rho^{\D-\mu-s}
\Bigl(\frac{-4}{x_{10}^2}\Bigr)^n  \\ 
 \sum_{k\ge0} \si^k \frac{\G(\D-s+k+n)}{k!} F \Bigl[ {\mu+k, \D-s+k+n
\atop \D-\mu-s+1} ;\rho \Bigr].
\end{multline}
Now we apply $m$ times the formula
\begin{align}
F \Bigl[{a,b \atop c}, z\Bigr] = F \Bigl[{a+1,b \atop c}, z\Bigr] -
\frac{b}{c} z F \Bigl[{a+1,b+1 \atop c+1}, z\Bigr],
\end{align}
see eq. 9.137,12 of \cite{GR}, and obtain a sum of series of
hypergeometric functions, of which each series can be summed with
9.182,8 of \cite{GR}, to give directly (\ref{Formel1}).

\section{Results for the propagators}

\subsection{The scalar case}

Now that we have all these formulae at hand, the calculation of the
scalar bulk-to-bulk propagator is ultra-simple. We convolute a
bulk-to-boundary propagator of dimension $\D$ with a bulk-to-boundary
propagator of dimension $\la$ ``along the boundary'', i.e. integrate
over the $d$ dimensional boundary variable and get with
(\ref{startofinduc}) for $s=0$
\begin{align}
A_{\la,\D} (x_1,x_3) & = \Phi_{0,0} (x_1,x_3) + \{\D \leftrightarrow
\la \}\nonumber \\
& = \frac{\pi^\mu \G(\mu-\D)} {\G(\la) \G(\D)} \La_{0,0} (\zeta) +
\{\D \leftrightarrow \la \}.
\end{align}
This is up to normalization just the AdS two-point function of a field
with parameter $\la$ plus the two-point function of a field with
parameter $\D$. Thus we see that in the computation of the AdS
two-point function by intertwining the legs of a conformal two-point
function we automatically obtain as an artefact of the technique the
two-point function of the field with the shadow parameter.


\subsection{The vector case}

Now we consider the convolution of two vector bulk-to-boundary
propagators, i.e. we consider the integral
\begin{align}
A^{(1)}_{\la,\D} (x_1, x_3) & = \int d^dx_2 K^{(1)}_\la (a, \vec
\nabla_b; x_{12}) K^{(1)}_\D (\vec b, c; x_{23}) \nonumber \\
& = \int d^dx_2 \frac{x_{10}^{\la-1}} {(x_{12}^2)^\la}
\Bigl( \frac{2 \langle a,x_{12} \rangle (\vec x_{12}, \vec \nabla_b)}
{x_{12}^2} -(\vec a, \vec \nabla_b) \Bigr) \nonumber \\
& \; \qquad \qquad \frac{x_{30}^{\D-1}} {(x_{23}^2)^\D}
\Bigl( \frac{2 (b, \vec x_{23}) \langle x_{23},c \rangle }
{x_{23}^2} -(\vec b, \vec c) \Bigr)
\end{align}
The idea of calculating this integral is to write each bulk to
boundary operator as a differential operator with respect to the
exterior variables acting on a linear combination of scalar
bulk-to-boundary like terms. We then get  
\begin{multline}
A^{(1)}_{\la,\D} (x_1, x_3) = \int d^dx_2 \biggl\{ -a_0
\frac{x_{10}^{\la}}{\la} (\vn_b, \vp_1) x_{12}^{-2\la} + \frac{1-\la}{\la}
x_{10}^{\la-1} (\vec a ,\vn_b) x_{12}^{-2\la} + \\
\frac{x_{10}^{\la-1}}{2 (\la-1)_2} (\vec a, \vp_1) (\vn_b, \vp_1)
x_{12}^{-2(\la-1)} \biggr\} \biggl\{ -c_0
\frac{x_{30}^{\D}}{\D} (\vec b, \vp_3) x_{23}^{-2\D} \\
+ \frac{1-\D}{\D} x_{30}^{\D-1} (\vec b ,\vec c) x_{23}^{-2\D} +
\frac{x_{30}^{\D-1}}{2 (\D-1)_2} (\vec b, \vp_3) (\vec c, \vp_3)
x_{12}^{-2(\la-1)} \biggr\}.
\end{multline}
Since all derivatives act on variables which are not integrated, we
can take them in front of the integrals and perform these in terms of
the functions $\Phi_{r,s}$, where we restrict on the direct term. The
$\Phi_{r,s}$ are functions of $x_{13}$, thus we can use $\vp:= \vp_1 =
-\vp_3$. The $\vec b$ are contracted and we are left with 
\begin{align}\label{A1inpowers}
A^{(1)}_{\la,\D} \Bigr \rvert_{\textrm{direct}} = A^{(1)}_{\la,\D}
\Bigr\rvert_{1} +  A^{(1)}_{\la,\D}
\Bigr\rvert_{a_0} + A^{(1)}_{\la,\D}
\Bigr\rvert_{c_0} + A^{(1)}_{\la,\D}\Bigr\rvert_{a_0 c_0},
\end{align}
where we expanded in powers of $a_0$ and $c_0$. The first term is
given by
\begin{align}
A^{(1)}_{\la,\D}& (x_1,x_3) \Bigr\rvert_{1} =  (\vec a, \vec c) \frac{(1-\la)
(1-\D)}{\la \D} \frac{1}{x_{10} x_{30}} \Phi_{0,0}(x_1,x_3) \nonumber \\ 
& + (\vec a,\vp)(\vec c,\vp) \biggl[ \frac{(\vp,\vp)} {2^2 (\la-1)_2
(\D-1)_2} \Phi_{1,1}(x_1,x_3) \nonumber\\
& + \frac{1}{x_{30}} \frac{1-\D}{2 (\la-1)_2 \D } \Phi_{1,0}(x_1,x_3)
+ \frac{1}{x_{10}} \frac{1-\la}{2 (\D-1)_2 \la } \Phi_{0,1}(x_1,x_3) \biggr]
\end{align}   
Now we observe that the first line can be presented directly as a Legendre
function by (\ref{Formel1}). For the term proportional to $(\vec
a,\vp)(\vec c,\vp)$ we use (\ref{LaplaceaufPhi}) on the first
summand in $[\cdots]$, and find that we can apply (\ref{Formel1}) on
the result
\begin{align}\label{resulvec1}
A^{(1)}_{\la,\D}&(x_1,x_3) \Bigr\rvert_{1}  =  (\vec a, \vec c)
\frac{(1-\la)(1-\D)}{\la \D} \frac{1}{x_{10} x_{30}}
\Phi_{0,0}(x_1,x_3) \nonumber \\  
 & \quad  + (\vec a,\vp)(\vec c,\vp) \biggl[ - \frac{1} {(\la-1)_2}
\Phi_{1,-1}(x_1,x_3) \nonumber \\
& \quad + \frac{1}{x_{30}} \frac{1-\la}{2 (\la-1)_2 \D }
\Phi_{1,0}(x_1,x_3) + \frac{1}{x_{10}} \frac{1-\la}{2 (\D-1)_2 \la }
\Phi_{0,1}(x_1,x_3) \biggr] \nonumber \\
& = \frac{\pi^\mu \G(\mu-\D)}{\G(\La+1) \G(\D+1)} \Biggl\{ - (\vec
a,\vp)(\vec c,\vp) \biggl[ \frac{\la-1}{2} 
\La_{1,1}(\zeta) + \frac{1}{\mu-\D-1} \La_{-1,0} (\zeta) \biggr] \nonumber \\
& \qquad\qquad\qquad\qquad\qquad + (\vec a, \vec
c)\frac{(1-\la)(1-\D)}{x_{10} x_{30}} \, \La_{0,0}(\zeta) \Biggr\}. 
\end{align}   

The second term in (\ref{A1inpowers}) is given by
\begin{align}
A^{(1)}_{\la,\D} \Bigr\rvert_{a_0} =  a_0 (\vec c,\vp) \biggl\{
-\frac{1}{2(\D-1)_2 \la} (\vp,\vp) \Phi_{0,1} - \frac{1}{x_{30}}
\frac{1-\D}{\la \D} \Phi_{0,0}\biggr\}.
\end{align}
We observe that both terms can be summed up by (\ref{Formel2}), to
result in
\begin{multline}\label{secofvecpart}
A^{(1)}_{\la,\D} (x_{10},x_{30}) \Bigr\rvert_{a_0} = \frac{\pi^\mu
\G(\mu-\D)}{ \G(\la+1) \G(\D+1)} a_0 (\vec c,\vp) \\ 
 \biggl\{ \frac{\la-1}{x_{30}} \,\La_{0,0}(\zeta) + \frac{1}{x_{30}}
\frac{2}{\mu-\D-1} \, \La_{-2,-1}(\zeta) + \frac{2}{x_{10}}  \,
\La_{-1,-1}(\zeta) \biggr\}.  
\end{multline}
In a similar manner we obtain for the third term in (\ref{A1inpowers})
\begin{align}\label{seaofvecpart}
A^{(1)}_{\la,\D}& (x_{10},x_{30}) \Bigr\rvert_{c_0} =  - c_0 (\vec
a,\vp) \biggl\{-\frac{1}{2(\la-1)_2 \D} (\vp,\vp) \Phi_{1,0} -
\frac{1}{x_{10}} \frac{1-\la}{\la \D} \Phi_{0,0}\biggr\} \nonumber \\ 
& =-\frac{\pi^\mu \G(\mu-\D)}{ \G(\la+1) \G(\D+1)} c_0 (\vec a,\vp)
\biggl\{ \frac{\la-1}{x_{10}} \, \La_{0,0}(\zeta) \nonumber \\ 
& \qquad\qquad\qquad  +\frac{1}{x_{10}} \frac{2}{\mu-\D-1} \,
\La_{-2,-1}(\zeta) + \frac{2}{x_{30}}  \, \La_{-1,-1}(\zeta)\biggr\}. 
\end{align}
We check that this one is the same as (\ref{secofvecpart}) after
interchanging $a \leftrightarrow c$ and $x_{1}\leftrightarrow x_{3}$,
up to the sign. This must be the case, because the derivative
$\vp=\vp_1$ acting on $\vec x_{13}$ equals $-\vp_3$, therefore the sum
of (\ref{secofvecpart}) and (\ref{seaofvecpart}) is symmetric under
the above permutation, which is evident from the defining
integral.\footnote{ Strictly speaking, it may happen that a possible
asymmetric part is cancelled with the asymmetric part in the shadow
term. However, from the comparison with
the vector propagator in the literature we know that this assumption
is indeed justified.} Finally, the fourth term in (\ref{A1inpowers})
is given by the Laplacian acting on a scalar bulk-to-bulk propagator
\begin{align}\label{resofvec2}
A^{(1)}_{\la,\D} (x_{10},x_{30}) \Bigr\rvert_{a_0 c_0} = \frac{1}{\la
\D} (\vp, \vp) \Phi_{0,0} =  \frac{\pi^\mu \G(\mu-\D)}{ \G(\la+1)
\G(\D+1)} a_0 c_0 (\vp, \vp) \, \La_{0,0}(\zeta).
\end{align} 

Let us now write the vector two-point function in terms of the
fundamental bitensors (\ref{fundbitens}). The two fundamental
bitensors for tensor rank $1$ are $\langle a, \partial
\zeta\rangle \langle c, \partial' \zeta \rangle$ and $\langle
a,\partial\rangle \langle c, \partial' \rangle \zeta$, where the
unprimed variables correspond to $x_1$ and the primed ones to
$x_3$. We perform the differentiations of the Legendre
functions $\La_{s,t}$ in
(\ref{resulvec1},\ref{secofvecpart},\ref{seaofvecpart},\ref{resofvec2})
with (\ref{AblvonLa}), e.g. let us look at (\ref{resulvec1})
\begin{align}\label{hauptvec}
A^{(1)}_{\la,\D}&(x_1,x_3) \Bigr\rvert_{1} = \frac{\pi^\mu
\G(\mu-\D)}{ \G(\la+1) \G(\D+1)}  \biggl\{ \nonumber \\
& \qquad\qquad 2 \Bigl[ (\la-1) \, \La_{-1,-1}(\zeta) +
\frac{2}{\mu-\D-1} \, \La_{-3,-2} (\zeta) \Bigr] (\vec a, \vp \zeta) (
\vec c, \vp' \zeta ) \nonumber \\
& \qquad\qquad + \Bigl[ \D (\la-1) \, \La_{0,0}(\zeta) +\frac{2}{\mu-\D-1} \,
\La_{-2,-1} (\zeta) \Bigr] \frac{(\vec a,\vec c)}{x_{10} x_{30}} \biggr\}, 
\end{align}
add the various contributions up, solve (\ref{covtononcov}) for the $d$
dimensional quantities, and insert them into the sum. We then obtain
for the complete vectorial two-point function 
\begin{align}\label{ergebvecprop} 
A^{(1)}_{\la,\D}&|_{direct} = \frac{\pi^\mu \G(\mu-\D)}{ \G(\la+1)
\G(\D+1)} \biggl\{2 \Bigl[ (\la-1) \, \La_{-1,-1}(\zeta) +
\frac{2}{\mu-\D-1} \, \La_{-3,-2} (\zeta) \Bigr] I_2 \nonumber \\
& \qquad\qquad\qquad\qquad - \Bigl[ \D (\la-1) \, \La_{0,0}(\zeta) +\frac{2}{\mu-\D-1} \,
\La_{-2,-1} (\zeta) \Bigr] I_1 \biggr\}. 
\end{align}
All other terms vanish, as can be checked with the identities
(\ref{Lambdaformeln}). Comparing (\ref{hauptvec}) with
(\ref{ergebvecprop}) one notes, that the result is completely
determined by the ``principal contribution'', i.e. terms of maximal
degree in $\vec a$ and $\vec c$. This is clear, because when
we substitute the $d$ dimensional expressions composed of $\vec a$ and
$\vec c$ by the $d+1$ dimensional
invariants, we add components with positive degree in $a_0$ or $c_0$,
which must cancel the contributions from (\ref{secofvecpart}-
\ref{resofvec2}), due to symmetry. Thus we only have to determine the
principal contributions.


\subsection{The symmetric traceless tensor of rank $2$}

In this case the complete calculations are unappealingly lengthy, but
with our experience gained from the vectorial case we know how to
shorten the route. In contrast to the case of the vector we now have to
take the subtraction of traces into account. First we write down the
bulk to boundary propagator
\begin{align}
 K^{(2)}_\la &(a, \vec b; x_{12}) =
\frac{x_{10}^{\la-2}}{(x_{12}^2)^{\la}} \biggl\{ \Bigl( \frac{2 \langle
a,x_{12} \rangle (\vec x_{12}, \vec b)} {x_{12}^2} -(\vec a,
\vec b) \Bigr)^2 - \frac{\vec b^2}{d}  \Bigl( \frac{2 \langle
a,x_{12} \rangle \vec x_{12}} {x_{12}^2} - \vec a \Bigr)^2
\biggr\}\nonumber \\
& = \frac{x_{10}^{\la-2}}{(x_{12}^2)^{\la}}\biggl\{ 4 \frac{(\vec a,\vec
x_{12})^2 (\vec b, \vec x_{12})^2} {x_{12}^4} - 4 \frac{(\vec a,\vec
b) (\vec a,\vec x_{12}) (\vec b, \vec x_{12})} {x_{12}^2}+ (\vec
a,\vec b)^2 \nonumber \\ 
&  \qquad  -\frac{\vec
b^2}{d} \Bigl(\vec a^2 - 4 x_{10}^2 \frac{(\vec a,\vec
x_{12})^2}{x_{12}^4} \Bigr) + \textrm{terms of
positive order in}\; a_0, c_0\biggr\}, 
\end{align}
which is traceless with respect to the bulk and the boundary
variables. As before we write down the integral for the bulk-to-bulk
propagator, restrict to the direct terms and expand into powers of
$a_0$ and $c_0$  
\begin{align}\label{Zerlina0c0}
2 A^{(2)}_{\la,\D}& (x_1, x_3) \Bigr\rvert_{\textrm{direct}}  = \int d^dx_2
K^{(2)}_\la (a, \vec \nabla_b; x_{12}) K^{(2)}_\D (\vec b, c; x_{23})
\Bigr\rvert_{\textrm{direct}} \nonumber \\ 
& = A^{(2)}_{\la,\D}(x_1, x_3)\Bigr\rvert_{p.c.} + \textrm{terms of
positive order in} \;a_0, c_0,
\end{align}
where $p.c.$ denotes the principal contribution, which is the only one
we have to take into account. The factor $2$ on the left hand side
comes from the two contractions of $\vn_b$ with $b$.

The calculation of $A^{(2)}_{\la,\D}\bigr\rvert_{p.c.}$ is similar to
the case of the vector propagator, i.e express it in terms of
derivatives of $\Phi_{r,s}$ and manipulate the resulting expressions with
the formula (\ref{LaplaceaufPhi}) until we can write the sum of
derivatives of the $\Phi_{r,s}$ functions in terms of the Legendre
functions $\La_{s,t}$ with (\ref{Formel1}) and (\ref{Formel2}).  We
decompose $A^{(2)}_{\la,\D}\bigr\rvert_{p.c.}$ into summands of
different numbers of derivatives: 
\begin{align} \label{A^(2)vier}
A^{(2)}_{\la,\D}\bigr\rvert_{p.c.} =& A^{(2)}_{\la,\D}
\Bigr\rvert_{(\vec a,\vp)^2 (\vec c,\vp)^2} +
A^{(2)}_{\la,\D}\Bigr\rvert_{(\vec a,\vp) (\vec c,\vp) (\vec a,\vec
c)} + A^{(2)}_{\la,\D} \Bigr\rvert_{(\vec a, \vec c)^2} \nonumber \\ 
& + A^{(2)}_{\la,\D} \Bigr\rvert_{(\vec a,\vp)^2 \vec c^2} +
A^{(2)}_{\la,\D} \Bigr\rvert_{\vec a^2 (\vec c,\vp)^2}  +
A^{(2)}_{\la,\D} \Bigr\rvert_{\vec a^2 \vec c^2}.
\end{align}
Before turning to the calculation, note that the invariant bitensors
we have to expect for the symmetric tensor of rank $2$ are $I_1^2, I_1
I_2, I_2^2$, and $I_3,I_4$ for the removal of the traces of the former
invariants. The terms in the second line of (\ref{A^(2)vier}) contain
$a^2$ and $c^2$ ( after adding appropriate terms with $a_0$ and
$c_0$), so they remove the traces of the terms in the first line and
can consequently be computed from them. Therefore it is sufficient to
consider only the terms in the first line.

Let us start with the calculation of the first term in (\ref{A^(2)vier}):
\begin{align}\label{beitr1vont2} 
A^{(2)}_{\la,\D}&(0,0;x_1,x_3) \Bigr\rvert_{(\vec a,\vp)^2 (\vec
c,\vp)^2} =  (\vec a,\vp)^2 (\vec c,\vp)^2 \biggl\{ \nonumber \\ 
& \quad \frac{2 (\vp,\vp)^2} {2^4 (\la-2)_4 (\D-2)_4} \Phi_{2,2} (x_1,x_3)
+ \frac{1}{x_{30}} \frac{(6-4\D)(\vp,\vp)} {2^3 (\la-2)_4 (\D-1)_3}
\Phi_{2,1} (x_1,x_3) \nonumber \\ 
& \quad + \frac{1}{x_{30}^2} \frac{(\D-1)_2} {2 (\la-2)_4 (\D)_2}
\Phi_{2,0} (x_1,x_3) + \frac{1}{x_{10}} \frac{(6-4\la)(\vp,\vp)} {2^3
(\la-1)_3 (\D-2)_4} \Phi_{1,2} (x_1,x_3) \nonumber \\ 
& \quad + \frac{(\la-2)(\D-2)-1+\mu} {(x_{10} x_{30})\,(\la-1)_3
(\D-1)_3} \Phi_{1,1} (x_1,x_3) + \frac{1}{x_{10}^2} \frac{(\la-1)_2}
{2 (\la)_2 (\D-2)_4} \Phi_{0,2} (x_1,x_3) \nonumber \\ 
& \quad + \frac{(\vp,\vp)} {2 d (\la-2)_4 (\D)_2} \Phi_{2,0} (x_1,x_3)
+ \frac{1}{x_{10}} \frac{d-2(\la-1)} {d(\la-1)_3 (\D)_2} \Phi_{1,0}
(x_1,x_3) \biggr\} \nonumber \\ 
& = 2 \frac{\pi^\mu \G(\mu-\D)}{ \G(\la+2) \G(\D+2)}(\vec a,\vp)^2 (\vec
c,\vp)^2
\biggl\{\frac{1}{(\mu-\D-2)_2} (1-\frac{1}{d}) \, \La_{-2,0}(\zeta)
\nonumber \\ 
& \qquad\qquad\qquad \quad \;\;+ \frac{(\la-1)}{\mu-\D-1}(1-\frac{1}{d}) \,
\La_{0,1}(\zeta) + \frac{1}{2^2} (\la-1)_2 \, \La_{2,2}(\zeta)
\biggr\}. 
\end{align}
The second term in (\ref{A^(2)vier}) is given by
\begin{align}\label{beitr2vont2} 
A^{(2)}_{\la,\D} & (x_1,x_3))\Bigr\rvert_{(\vec a,\vp) (\vec c,\vp) (\vec a,
\vec c)} = (\vec a,\vp) (\vec c,\vp) (\vec a,
\vec c) \biggl\{  \nonumber \\
& \!\!\!- \frac{1}{x_{10} x_{30}^2} \frac{2 (\la-1)(\D-1)_2}{(\la-1)_3
 (\D)_2} \Phi_{1,0}(x_1,x_3) -
\frac{1}{x_{10}^2 x_{30}} \frac{2 (\la-1)_2 (\D-1)}{(\la)_2 (\D-1)_3} \Phi_{0,1
}(x_1,x_3)  \nonumber\\
& \qquad \qquad \qquad \qquad \qquad \qquad + \frac{1}{x_{10} x_{30}} \frac{(\la-1)(\D-1)}{(\la-1)_3
(\D-1)_3}  (\vp,\vp) \Phi_{1,1}(x_1,x_3) \biggr\} \nonumber \\
& \!\!\!\!\!\! = 2 \frac{\pi^\mu \G(\mu-\D)}{ \G(\la+2) \G(\D+2)}(\vec a,\vp) (\vec
c,\vp) (\vec a, \vec c) \nonumber\\
& \qquad \qquad \qquad \qquad\quad \frac{1-\D}{x_{10} x_{30}} \biggl\{ (\la-1)_2 \La_{1,1}(\zeta)+ 2
\frac{\la-1}{\mu-\D-1} \La_{-1,0}(\zeta) \biggr\}.
\end{align}
Finally, the third summand in (\ref{A^(2)vier}) is
\begin{align}\label{beitr3vont2} 
A^{(2)}_{\la,\D}&(x_1,x_3)\Bigr\rvert_{(\vec a, \vec c)^2} = (\vec a, \vec c)^2
\frac{2}{x_{10}^2 x_{30}^2} \frac{(\la-1)_2 (\D-1)_2} {(\la)_2 (\D)_2}
\Phi_{0,0}(x_1,x_3) \nonumber \\
&  = 2 \frac{\pi^\mu \G(\mu-\D)}{ \G(\la+2) \G(\D+2)} (\vec a, \vec c)^2 \frac{(\la-1)_2
(\D-1)_2}{x_{10}^2 x_{30}^2} \La_{0,0}(\zeta).
\end{align}
Now it is straightforward to write down the full propagator for a
symmetric traceless tensor field of rank $2$. We perform the
differentiations in (\ref{beitr1vont2}) and (\ref{beitr2vont2}), add
them up together with (\ref{beitr3vont2}) and sort them with respect
to the invariants $I_1^2, I_2^2$, and $I_1 I_2$. After some not very
labourious algebra we then find the propagator of a symmetric traceless
tensor field of rank $2$
\begin{align}
A^{(2)}_{\la,\D} & (x_1, x_3) \Bigr\rvert_{\textrm{direct}} =
\frac{\pi^\mu \G(\mu-\D)}{ \G(\la+2) \G(\D+2)} \biggl\{ \nonumber \\
& \quad\, \Bigl[2^4 \frac{d-1}{d} \, \La_{-4,-4}(\zeta) + 2^3
\frac{d-1}{\mu-\D-1} \, \La_{-4,-3}(\zeta) + 2^2 (\la-1)_2 \,
\La_{-2,-2}(\zeta) \Bigr] I_2^2 \nonumber \\
& + \Bigl[-2^5 \frac{d-1}{d} \, \La_{-3,-3}(\zeta) - 2^3
\frac{\la \D+d-1}{\mu-\D-1} \, \La_{-3,-2}(\zeta) \nonumber \\
& \qquad \qquad \qquad \qquad\qquad \qquad \qquad\qquad  - 2^2
(\D+1)(\la-1)_2\, \La_{-1,-1}(\zeta) \Bigr] I_1 I_2 \nonumber \\
& + \Bigl[ 2^3 \frac{d-1}{d} \, \La_{-2,-2}(\zeta) + 2^2
\frac{\la \D}{\mu-\D-1} \, \La_{-2,-1}(\zeta) + (\D)_2 (\la-1)_2 \,
\La_{0,0}(\zeta) \Bigr] I_1^2 \nonumber \\
& - \textrm{traces}\biggr\}.
\end{align}

\subsection*{Acknowledgements}

This work is supported in part by the German Volkswagenstiftung. The
work of R.~M. was supported by DFG (Deutsche Forschungsgemeinschaft).

\section*{Appendix: The trace terms}

Consistency of the group approach to traceless symmetric tensor fields
on AdS space requires that the bitensor propagators can be made
traceless using only the basis of geometric bitensors $I_1,I_2,I_3$
and $I_4$ from (\ref{Invars1},\ref{Invars2}). We apply the Laplacian
$\D_a$ to equation (\ref{AnsFuerSpurLos}) and express the result as a
linear combination of the independent monomials
\begin{align}
I_1^{m_1'} I_2^{m_2'}I_3^{n_1'}I_4^{n_1'} R_{1,2}
\end{align}
with
\begin{align}
R_1 := \frac{1}{z_0^2} I_{2c}^2, \quad R_2 := \frac{1}{a^2} I_4
\end{align}
In fact, the basic formula is 
\begin{align}\label{basicformula}
\frac{1}{2} & \D_a I_1^{m_1} I_2^{m_2} I_3^{n_1} I_4^{n_2} = \Bigl({m_1
\atop 2} \Bigr) (R_1 +R_2)  I_1^{m_1-2} I_2^{m_2} I_3^{n_1} I_4^{n_2}
\nonumber \\
& + m_1 m_2 \zeta R_1  I_1^{m_1-1} I_2^{m_2-1} I_3^{n_1} I_4^{n_2} +
\Bigl({m_2 \atop 2}\Bigr) (\zeta^2-1) R_1  I_1^{m_1} I_2^{m_2-2}
I_3^{n_1} I_4^{n_2} \nonumber \\
& + \biggl[ 4 \Bigl({n_1 \atop 2} \Bigr) \bigl(R_1 +(\zeta^2-1) R_2
\bigr) + n_1 \bigl((d+1) R_1 + (\zeta^2-1) R_2 \bigr) \nonumber \\
& \qquad \qquad \qquad + 2 m_1 n_1 R_1 
+ 2 m_2 n_1 \bigl( R_1 +(\zeta^2-1) R_2 \bigr) \biggr] I_1^{m_1}
I_2^{m_2} I_3^{n_1-1} I_4^{n_2} \nonumber \\
& + \biggl[ 4 \Bigl({n_2 \atop 2} \Bigr) R_2 + n_2 (d+1) R_2 +
2(m_1+m_2+2 n_1)n_2 R_2 \biggr] I_1^{m_1}
I_2^{m_2} I_3^{n_1} I_4^{n_2-1} \nonumber \\
& + 4 \Bigl({n_1 \atop 2} \Bigr) R_2 I_1^{m_1} I_2^{m_2+2} I_3^{n_1-2}
I_4^{n_2} - 4 \Bigl({n_1 \atop 2} \Bigr) (\zeta^2-1) R_1 I_1^{m_1}
I_2^{m_2} I_3^{n_1-2} I_4^{n_2+1} \nonumber \\
& + 2 m_1 n_1 \zeta R_2 I_1^{m_1-1} I_2^{m_2+1} I_3^{n_1-1} I_4^{n_2}
\end{align}
In the case $l=2$, we have $(l_1,l_2) \in \{(2,0), (1,1), (0,2) \}$
and we calculate the matrix $\mathcal{V}$ representing $\frac{1}{2}
\D_a$ in the bases $\{I_1^{l_1} I_2^{l_2}\}$ and $\{R_1,R_2\}$
respectively. We obtain
\begin{align}
\begin{array}{c c}
 & \begin{array} {c c c} \; (2,0) \; & \;(1,1)\; & \;(0,2)\; \end{array} \\
\begin{array} {c} R_1 \\ R_2 \end{array} &
 \left[ \begin{array}{c c c} \quad 1 \quad\; & \quad \zeta \quad & \;\zeta^2-1 \\\quad 1\quad\; & 0 & \;0 \end{array} \right]
\end{array}
\end{align}
where we indicated the respective basis elements by the labels above each
column to the left of the rows. For the second term in
(\ref{AnsFuerSpurLos}) we need the matrix $\mathcal{M}$ representing
$\frac{1}{2}\D_a$ on $span(I_3,I_4)$, and the image is again
$span(R_1,R_2)$. This must be the case, because otherwise there were
no solution to (\ref{AnsFuerSpurLos}). We get from (\ref{basicformula})
\begin{align}
\mathcal{M} = \left[ \begin{array}{c c} d+1 & 0 \\ \zeta^2-1 & d+1
 \end{array}\right]
\end{align}
and collect the unknown coefficients in a matrix
\begin{align}
\mathcal{A} = \left[\begin{array}{c c c}  A^{(2,0)}_{0010} &
A^{(1,1)}_{0010} & A^{(0,2)}_{0010} \\ A^{(2,0)}_{0001} &
A^{(1,1)}_{0001} & A^{(0,2)}_{0001} \end{array} \right].
\end{align}
Then (\ref{AnsFuerSpurLos}) takes the form of a matrix equation 
\begin{align}
\mathcal{V} = \mathcal{M A},
\end{align}
which is easy to solve thanks to the triangular shape of
$\mathcal{M}$:
\begin{align}
\mathcal{A} = \mathcal{M}^{-1} \mathcal{V} = \frac{1}{d+1}
\left[ \begin{array}{c c c}  1 & \zeta & \zeta^2-1 \\
1-\frac{\zeta^2-1}{d+1} & -\zeta \frac{\zeta^2-1}{d+1} & -
\frac{(\zeta^2-1)^2}{d+1} \end{array}\right].
\end{align} 

In the case $l=3$ we also get a triangular $4 \times 4$ matrix
$\mathcal{M}$ acting on the trace terms and the matrix $\mathcal{A}$
of unknowns comes out as
\begin{multline}
\left[ \begin{array} {c c c c} A^{(30)}_{1010} & A^{(21)}_{1010} &
A^{(12)}_{1010} & A^{(03)}_{1010} \\
A^{(30)}_{1001} & A^{(21)}_{1001} & A^{(12)}_{1001} & A^{(03)}_{1001}
\\  
A^{(30)}_{0110} & A^{(21)}_{0110} & A^{(12)}_{0110} & A^{(03)}_{0110}
\\
A^{(30)}_{0101} & A^{(21)}_{0101} & A^{(12)}_{0101} & A^{(03)}_{0101}
\end{array}\right] \\ = \frac{1}{d+3} \left[ \begin{array}{c c c c}
3 & 2\zeta & \zeta^2-1 & 0 \\ 3 \bigl(1- \frac{\zeta^2-1}{d+3} \bigr)
& - 2 \zeta\frac{\zeta^2-1}{d+3} & - \frac{(\zeta^2-1)^2}{d+3} & 0 \\
0 & 1 & 2 \zeta & 3(\zeta^2-1) \\ - \frac{6 \zeta}{d+3} & 1 -
\frac{7\zeta^2-3}{d+3} & - 8 \zeta \frac{\zeta^2-1}{d+3} & -9
\frac{(\zeta^2-1)^2}{d+3} \end{array}\right].
\end{multline}

In the case $l=4$ a dramatic change occurs since the matrix
$\mathcal{M}$ representing $\frac{1}{2} \D_a$ on the rank $4$ trace
terms becomes a $9 \times 10$ rectangular matrix, i.e. there are $9$
trace terms to cancel the traces of the $10$ invariants of rank
$4$, thereby leading to 10 equations for the unknown entries of
$\mathcal{A}$. In general, for $l=2p$ even, the number of trace terms and the
number of invariants start to differ at $p\ge 2$:
\begin{align}
\textrm{Number of trace terms} & = p \bigl(\frac{1}{3} p^2 + \frac{3}{2}
p +\frac{1}{6} \bigr) \nonumber \\
\textrm{Number of equations} & = p (p+1) \bigl( \frac{2}{3} p
+\frac{1}{3} \bigr).
\end{align}
Thus the set of equations for $\mathcal{A}$ contains 
\begin{align}
p \bigl( \frac{1}{3} p^2 - \frac{1}{2} p +\frac{1}{6} \bigr)
\end{align}
linearly dependent equations. We compute the linear system for $l=4$
and show thereby that there is $1$ linearly dependent equation. The
matrix $\mathcal{V}$ representing $\frac{1}{2} \D_a$ on
\begin{align}\label{basisorderpreim}
span\{I_1^{4}, I_1^3 I_2, I_1^2 I_2^2, I_1 I_2^3, I_2^4 \}
\end{align}
is calculated by (\ref{basicformula}) to result in a $10 \times 5$
matrix, which decomposes into 
\begin{align}
\mathcal{V} = \left[ \begin{array}{c} \mathcal{V}_0 \\ 0 \end{array}
\right],
\end{align}
where
\begin{align}
\mathcal{V}_0 = \left[ \begin{array}{c c c c c} 
6 & 3 \zeta & \zeta^2-1 & 0 & 0 \\
6 & 0 & 0 & 0 & 0 \\
0 & 3 & 4\zeta & 3 (\zeta^2-1) & 0 \\
0 & 3 & 0 & 0 & 0 \\
0 & 0 & 1 & 3 \zeta & \zeta^2 -1 \\
0 & 0 & 1 & 0 & 0 \end{array} \right].
\end{align}
The basis in the preimage of $\frac{1}{2} \D_a$ is ordered as in
(\ref{basisorderpreim}), and the $10$ basis elements $\{I_i I_j R_k\}$
in the image are ordered lexicographically, i.e. $I_1^2 R_1, I_1^2
R_2, \ldots, I_4 R_2$. By the simple transformation 
\begin{align}\label{simpletransf}
A^{(l_1 l_2)}_{0201} \mapsto A^{(l_1 l_2)}_{0201} + \frac{4}{d+5}
A_{0020}^{(l_1 l_2)} 
\end{align}
we can bring the system of equations into the block form
\begin{align}
\left[\begin{array}{c} \mathcal{V}_0 \\ 0 \end{array} \right] = \left[
\begin{array}{c c} \mathcal{M}_0 & 0 \\ \mathcal{B} & \mathcal{M}_1
\end{array} \right] \left[\begin{array}{c} \mathcal{A}_0 \\
\mathcal{A}_1 \end{array} \right],
\end{align}
where the $4$-tupels $(m_1, m_2, n_1, n_2)$ are ordered as 
\begin{align}
(2010),(2001), (1110), (1101), (0210), (0201), (0020), (0011), (0002).
\end{align}
Then $\mathcal{M}_0$ has triangular shape 
\begin{align}
\mathcal{M}_0 =\left[ \begin{array}{c c c c c c}
d+5 & 0 & 0 & 0 & 0 & 0 \\
\zeta^2-1 & d+5 & 0 & 0 & 0 & 0 \\
0 & 0 & d+5 & 0 & 0 & 0 \\
4\zeta & 0 & 3 (\zeta^2-1) & d+5 & 0 & 0 \\
0 & 0 & 0 & 0 & d+5 & 0 \\
0 & 0 & 2 \zeta & 0 & 5(\zeta^2-1) & d+5 \end{array} \right],
\end{align}
thus $\mathcal{M}_0$ can be inverted easily to give
\begin{align}
\mathcal{A}_0 = \frac{1}{d+5} \left[ \begin{array} {c c c c c} 
6 & 3 \zeta & \zeta^2-1 & 0 & 0 \\
6 \bigl(1-\frac{\zeta^2-1}{d+5} \bigr) & - 3\zeta
\frac{\zeta^2-1}{d+5} & -\frac{(\zeta^2-1)^2}{d+5} & 0 & 0 \\
0 & 3 & 4\zeta & 3 (\zeta^2-1) & 0 \\
-\frac{24 \zeta}{d+5} & 3 \bigl(1-\frac{7 \zeta^2-3}{d+5} \bigr) & -
16 \zeta \frac{(\zeta^2-1)^2}{d+5} & -9 \frac{(\zeta^2-1)^2}{d+5} & 0
\\
0 & 0 & 1 & 3 \zeta & 6(\zeta^2-1) \\
0 & - \frac{6 \zeta}{d+5} & 1-\frac{13 \zeta^2-5}{d+5} & -21 \zeta
\frac{\zeta^2-1}{d+5} & -30 \frac{(\zeta^2-1)^2}{d+5} \end{array}
\right]
\end{align} 

Finally it remains to solve
\begin{align}\label{remaintosolve}
- \mathcal{B A}_0 = \mathcal{M}_1 \mathcal{A}_1.
\end{align}
We manipulate both sides the following way: Multiplying the first row
with $\frac{\zeta (\zeta^2-1)}{d+5}$, the second with
$-1+\frac{4}{d+5}$ and adding both to the third row results in a zero
row, which we skip on both sides of (\ref{remaintosolve}). We thus get
the transformed equations
\begin{align}
\mathcal{M}_1 \mapsto \widetilde{\mathcal{M}}_1, \quad \mathcal{B A}_0
\mapsto \widetilde{ \mathcal{B A}}_0
\end{align}
with 
\begin{align}
\widetilde{ \mathcal{M}}_1 = \left[ \begin{array}{c c c} 2(d+3) & 0 & 0 \\ 6
(\zeta^2 -1) & d+5 & 0 \\ 0 & (\zeta^2-1) & 2(d+3) \end{array}
\right]
\end{align}
and
\begin{align}
\widetilde{ \mathcal{B A}}_0 = \frac{1}{d+5} \left[ \begin{array}{c c c c c}
6 & 6 \zeta & 2(3\zeta^2-1) & 6 \zeta(\zeta^2-1) & 6 (\zeta^2-1)^2 \\
6 & 3 \zeta & \zeta^2-1 & 0 & 0 \\
6 \bigl( 1-\frac{\zeta^2-1}{d+5} \bigr) & -3 \zeta
\frac{\zeta^2-1}{d+5} & - \frac{(\zeta^2-1)^2}{d+5} & 0 & 0
\end{array} \right]
\end{align}
We evaluate then 
\begin{align}
\mathcal{A}_1 = -\widetilde{ \mathcal{M}}_1^{-1} \widetilde{ \mathcal{B A}}_0 
\end{align}
and get
\begin{multline}
\mathcal{A}_1 = -\frac{1}{d+5} 
\left[ \begin{array}{c c}
\frac{3}{d+3} & \frac{3\zeta}{d+3} \\ 
\frac{6}{d+5} \bigl(1 -\frac{3 (\zeta^2-1)}{d+3} \bigr) & \frac{3
\zeta}{d+5} \bigl(1 -\frac{6 (\zeta^2-1)}{d+3} \bigr) \\
\frac{3}{d+3}\bigl( 1 - \frac{2(\zeta^2-1)}{d+5} + \frac{3
(\zeta^2-1)^2}{(d+3)(d+5)} \bigr) & \frac{3 \zeta (\zeta^2-1)}{(d+3)(d+5)}
\bigl(-1+\frac{3(\zeta^2-1)}{d+3} \bigr) \end{array} \right. \\
\left. \begin{array}{c c c} \frac{3\zeta^2-1}{d+3} &
\frac{3\zeta (\zeta^2-1)}{d+3} & \frac{3 (\zeta^2-1)^2}{d+3} \\
\frac{\zeta^2-1}{d+5} \bigl( 1-\frac{6(3\zeta^2-1)}{d+3} \Bigr) & \frac{-18 \zeta
(\zeta^2-1)^2}{(d+3)(d+5)} & \frac{-18 (\zeta^2-1)^3}{(d+3)(d+5)} \\
\frac{(\zeta^2-1)^2}{(d+3)(d+5)}\bigl(-1+\frac{3(3\zeta^2-1)}{d+3}
\bigr) & \frac{9 \zeta(\zeta^2-1)^3}{(d+3)^2(d+5)} & \frac{9
(\zeta^2-1)^4}{(d+3)^2 (d+5)} \end{array} \right]
\end{multline}
At the end, we have to invert the transformation (\ref{simpletransf}).

\end{document}